\documentstyle[12pt,aaspp]{article}

\def\ltsima{$\; \buildrel < \over \sim \;$}
\def\lsim{\lower.5ex\hbox{\ltsima}}
\def\gtsima{$\; \buildrel > \over \sim \;$}
\def\gsim{\lower.5ex\hbox{\gtsima}}

\begin{document}
\title
{AN X-RAY MICROLENSING TEST OF 
AU-SCALE ACCRETION DISK STRUCTURE IN Q2237+0305}

\author{Atsunori Yonehara\altaffilmark{1,2}, Shin Mineshige\altaffilmark{1},
Tadahiro Manmoto\altaffilmark{1},}

\author{Jun Fukue\altaffilmark{3}, Masayuki Umemura\altaffilmark{4},
Edwin L. Turner\altaffilmark{5}}

\altaffiltext{1}{Department of Astronomy, Kyoto University, Sakyo-ku, 
 Kyoto 606-8502, Japan}
\altaffiltext{2}{e-mail: yonehara@kusastro.kyoto-u.ac.jp}
\altaffiltext{3}{Astronomical Institute, Osaka Kyoiku University,
Asahigaoka, Kashiwara, Osaka 582-0026, Japan}
\altaffiltext{4}{Center for Computational Physics, Tsukuba University,
Tsukuba, Ibaraki 305-0006, Japan}
\altaffiltext{5}{Princeton University Observatory, Peyton Hall, Princeton, NJ
08544, USA}

\begin{abstract}

The innermost regions of quasars can be resolved by a gravitational-lens 
{\lq}telescope{\rq} on scales down to a few AU.
For the purpose, X-ray observations are most preferable, 
because X-rays originating from the innermost regions,   
can be selectively amplified by microlensing 
due to the so-called `caustic crossing'.
If detected, X-ray variations will constrain 
the size of the X-ray emitting region down to a few AU.
The maximum attainable resolution depends mainly on
the monitoring intervals of lens events, 
which should be much shorter than the crossing time.
On the basis of this idea, 
we performe numerical simulations of microlensing of 
an optically-thick, standard-type disk as well as 
an optically-thin, advection-dominated accretion flow (ADAF). 
Calculated spectral variations and light curves show distinct
behaviors, depending on the photon energy. 
X-ray radiation which is produced in optically thin region, 
exhibits intensity variation over a few tens of days. 
In contrast, optical-UV fluxes, which are likely to 
come from optically thick region, 
exhibit more gradual light changes, which is consistent with 
the microlensing events so far observed in Q2237+0305. 

Currently, Q2237+0305 is being monitored in the optical range 
at Apache Point Observatory. 
Simultaneous multi-wavelength observations by X-ray sattelites 
(e.g., ASCA, AXAF, XMM) as well as HST at the moment of 
a microlens event enable us to reveal 
an AU scale structure of the central accretion disk around black hole.

\end{abstract}

\keywords{accretion, accretion disks --- active galactic nuclei
 --- microlensing --- quasars: individuals (Q2237+0305)}

\section{INTRODUCTION}

The high power output from quasars is usually attributed to 
the combination of a supermassive black hole with a surrounding accretion disk.
This belief is supported by a number of observations that indicate the
presence of a deep gravitational potential well or a hot gas disk 
at the center of quasars or other active galactic nuclei;
e.g., measurements of stellar velocity dispersion 
clearly showed a peculiar increase toward the center 
(Young et al. 1978; Sargent et al. 1978; 
see also Ford et al. 1994; Harms et al. 1994).  
Malkan (1983) found that the optical to UV spectra 
are well fitted by the standard-type accretion disk model
(Shakura \& Sunyaev 1973).
Recently, by far the best evidence of a supermassive black hole has been found
by radio observations of nuclear H$_2$O maser 
sources in NGC4258 (Miyoshi et al. 1995).
Alternatively, we can infer the presence of a relativistic object from
the asymmetric Fe line profile (Tanaka et al. 1995).
These observational results are all attractive, but still the real vicinity of 
a putative black hole has not been resolved. 

Q2237+0305 (e.g., Huchra et al. 1985) is the first object, 
in which the quasar microlensing events were detected 
(Corrigan et al. 1987; Houde \& Racine 1994; see also Ostensen et al. 1996).
These observations suggest that microlensing events 
take place roughly once per year.
This rather high frequency is consistent with the microlens  
optical depth of $\tau \sim 0.8$ obtained by the realistic simulation 
of the lensing galaxy (i.e., Wambsganss \& Paczy\'nski 1994).
We consider, here, specifically the microlensing events of this source 
caused by the so-called `caustic crossings' 
(see Yonehara et al. 1997 for single-lens calculations).
Several authors have already analyzed this `caustic' case 
based on a simple model for quasar accretion disk 
(e.g., Wambsganss \& Paczy\'nski 1991; 
Jaroszy\'nski, Wambsganss \& Paczy\'nski 1992). 
So far, however, only the standard-type disk, 
which is too cool to emit X-rays, has been considered, 
and thus the property of an X-ray microlensing of quasar, e.g., Q2237+0305,  
has not been predicted. 
We stress here the significance of X-ray observations to elucidate the physics 
of the innermost parts of the disk, since X-rays specifically originate 
from a deep potential well. 
The observations allow us to assess the extension of hot regions on 
several AU scales and resultantly to deduce 
the mass of a central massive black hole.

In this $Letter$, we propose to investigate quasar central structure 
by using X-ray microlensing of Q2237+0305. 
In section 2, we describe the method for resolving X-ray emission properties 
of the inner disk structure on a scale down to a few AU. 
In section 3, we calculate the disk spectra 
and light curves during microlensing. 
We here use realistic disk models: 
the optically-thick, standard disk (Shakura \& Sunyaev 1973) and 
the optically-thin, advection-dominated accretion flow 
(ADAF, Abramowicz et al. 1995; Narayan \& Yi 1995; see also Ichimaru 1977).

\section{X-RAY MICROLENS}

Using microlensing, it is possible to discriminate the 
structure of quasar accretion disks, and possibly to map them in detail.
Broad band photometry will be able to detect the color changes, 
thereby revealing the structure of quasar accretion disks.
Fortunately, such observations do not require very high  
time resolution, if the lens is far from the observer.
In the case of Q2237+0305, furthermore, 
all the four images are very close, $\sim 1 {\rm arcsec}$, 
to the image center of the lensing galaxy (e.g., Irwin et al. 1989).
This situation and the almost symmetric image pattern  
permits us to neglect time delay between images; 
e.g., time delay between images A and B is $\sim 3 {\rm hr}$ 
for $H_{\rm 0}=75 {\rm km~s^{-1} Mpc^{-1}}$ (Wambsganss \& Paczy\'nski 1994). 
We can thus easily discriminate intrinsic variability from microlensing;  
if only one image exhibits peculiar brightening over several tens of days 
superposed on intrinsic variations which the other three also show, 
we can conclude that the event is due to microlensing.  
On the other hand, variations caused by 
superposition of other variable objects 
(e.g., supernovae and cataclysmic variables) are distinguishable 
in terms of their characteristic light curves shapes 
(e.g., sharp rise, reccurency, etc.).

One important length scale for lensing is the {\it Einstein-ring radius}  
on the source plane, $r_{\rm E} \equiv \theta_{\rm E}D_{\rm os}$, where 
$\theta_{\rm E}=[(4GM_{\rm lens}/c^2)(D_{\rm ls}/D_{\rm os}D_{\rm ol})]^{1/2}$,
$M_{\rm lens}$ is the typical mass of a lens star, 
and $D_{\rm ls}$, $D_{\rm os}$, and $D_{\rm ol}$ 
represent the angular diameter distances from the lens to the source, 
from the observer to the source, and from the observer to the lens, 
respectively (see Paczy\'nski 1986). For Q2237+0305 (e.g., Irwin 1989), 
the redshifts corresponding to the distances from the observer to the quasar,
from the observer to the lens, and from the lens to the quasar are,
$z_{\rm os} = 1.675$, $z_{\rm ol} = 0.039$,
and $z_{\rm ls} = 1.575$, respectively.
For these the Einstein-ring angular extension 
and the radius on the source plane are 
$\theta_{\rm E} \sim 4 \times 10^{-11} (M_{\rm lens}/M_{\odot}) {\rm (radian)}$
 and $r_{\rm E} \equiv \theta_{\rm E}D_{\rm os} 
\sim 1.5\times 10^{17}(M_{\rm lens}/M_\odot)^{1/2}$cm, respectively, 
where we assume an Einstein-de Sitter universe and
Hubble's constant to be $H_{\rm 0} \sim 60 {\rm km \ s^{-1} Mpc^{-1}}$, 
according to Kundi\'c et al. (1997).

We, however, emphasize that an even more important length is 
the {\it caustic crossing scale} on the quasar image plane ($r_{\rm cross}$);
\begin{equation}
 r_{\rm cross} 
    = v_{\rm t}t \frac{D_{\rm os}}{D_{\rm ol}} 
           \sim 4.1\times 10^{13}
               \left(\frac{v_{\rm t}}{600~{\rm km~s}^{-1}}\right)
               \left(\frac{t}{1~{\rm d}}\right) {\rm cm},
\end{equation}
where $t$ is the time over which the caustic moves on 
the quasar disk plane and $v_{\rm t}$ is the transverse velocity of the lens, 
including the transverse velocity of the peculiar motion of 
the foreground galaxy relative to the source and the observer.
Surprisingly, this is comparable to the Schwarzschild radius,
$r_{\rm g}\simeq 3\times 10^{13}M_8$cm,
for a $10^8M_8$ black hole and is much smaller than $r_{\rm E}$.  
In other words, by daily monitoring, we can hope to resolve the disk structure 
with a resolution of $\sim$ 2.7 AU or $\sim r_{\rm g}M_8^{-1}$, 
if a microlensing event takes place.

We can then place a limit on the depth of the potential well 
at the center of a quasar
from X-ray observations of the microlensing light curve.
Since an X-ray emitting plasma is within the potential well, 
we can conjecture 
$kT_{\rm e}/m_{\rm p} < GM/r$ (with $k$ being the Boltzmann constant, 
$T_{\rm e}$ being electron temperature 
and $m_{\rm p}$ being the proton mass); 
This leads to a constraint for the central hole mass as  
\begin{equation}
  M > 1.5 \times 10^5 M_\odot 
                 \left(\frac{r_{\rm cross}}{2.5\times 10^{14}{\rm cm}}\right)
                 \left(\frac{T_{\rm e}}{10^9{\rm K}}\right).
\end{equation}
If we observe every 6 days, 
we can achieve an effective spatial resolution 
of $2.5\times 10^{14}$cm on the disk plane and thus set a lower mass limit
of $\sim 10^{5}M_\odot$ for $T_{\rm e} \sim 10^9{\rm K}$ 
in such a compact region. 
The maximally attainable resolution is, therefore, 
over two orders of magnitude better than 
what was obtained by ${\rm H_2O}$ maser observation. 

Although X-ray observations alone are very informative,  
by simultaneous X-ray and optical observations 
allow us to obtain more detailed information 
concerning AU-scale disk structure. 
We can set good constrain to the model parameters, 
such as $v_{\rm t}$ from optical observations, 
because the standard-type disk, emitting the optical flux, is well studied. 
We demonstrate this bellow (\S 3.3).

\section{MICROLENS DIAGNOSIS}

\subsection{Simulation Methods}

Let us next calculate expected X-ray variations based on specific disk models. 
The observed flux from a part of the disk at ($r_i,\varphi_j$) 
during a microlensing event is calculated by
\begin{equation}
\Delta F_{\rm obs}(\nu;r_i,\varphi_j) d \Omega \simeq 
A(u) \frac{\Delta L[\nu(1+z_{\rm os}),r_i]}{4\pi D_{\rm os}^2(1+z_{\rm os})^3} 
r_i dr_i \varphi_j .
\end{equation}
The integral $\int\Delta F_{\rm obs} d\Omega$ over the disk plane gives
the total observed flux.
Here, $\Delta L$ is the luminosity of a part of a disk, 
$A(u)$ represents the amplification factor as a function of
the angular separation ($u$) between the source and a caustic, and 
$z_{\rm os}$ is the redshift between the observer and the source.  
The amplification factor is calculated in the following way. 
We consider the idealized situation of a straight caustic
with infinite length passing over the disk.
This is a crude but reasonable approximation,
since the size of the source (i.e., disk) of interest is smaller 
than the Einstein-ring radius on the source plane.
Then, the following analytical 
approximate formula can be used (see Schneider, Ehlers, \& Falco, 1992): 
\begin{equation}
A(u) = \left\{
	\begin{array}{@{\,}ll}
	 u^{-1/2} + A_{\rm 0} & \mbox{(\rm for $u > 0$)} \\
	 A_{\rm 0} & \mbox{(\rm for $u \le 0$)},
	\end{array}
       \right. 
\label{eq:ampli}
\end{equation}
where $u$ is the angular separation from a caustic in units of 
$\theta_{\rm E}$, and $A_{\rm 0}$ represents a constant amplification factor 
due to the initial amplification by the caustic  
(i.e., when the source lies outside the caustic) and 
also due to the effects of other caustics. 
In the present study, we set $A_{\rm 0}=6.0$ so as to reproduce 
observed microlens amplitudes $\sim 0.5 {\rm mag}$. 
The positive (or negative) angular separation, $u>0$ ($u<0$), means that 
the source is located inside (outside) the caustic (see figure 1).

To calculate $\Delta L$, we consider a hybrid disk model 
which is composed of an optically thick standard disk 
and an optically thin ADAF. 
Optical-UV flux come from optically thick parts 
while X-rays originate from optically thin ADAF parts.
For the optically thick part, we assume blackbody radiation
$B_{\nu}(T)$ (with $\nu$ being frequency), 
where temperature $T(r)$ is given by
\begin{equation}
T(r) = 2.2 \times 10^5 \left( \frac{\dot{M}}{10^{26}
 {\rm g \ s^{-1}}} \right)^{1/4} \left( \frac{M}{10^8 M_{\odot}}
\right)^{1/4} \left( \frac{r}{10^{14} {\rm cm}} \right)^{-3/4}
\left[1 - \left( \frac{r_{\rm in}}{r} \right)^{1/2} \right]^{1/4} {\rm K},
\end{equation}
if relativistic effects, irradiation heating, 
 and Compton scattering effects are ignored.
Here, $\dot{M}$ is the mass accretion rate, $M$ is the black hole mass, and 
the inner edge of the disk ($r_{\rm in}$) is set to be $3r_{\rm g}$.
The fractional luminosity from a tiny portion of the disk at $(r,\phi)$
with surface element $\Delta S =r\Delta r\Delta\phi$ is then
$\Delta L(\nu,r) = 8\pi B_{\nu}\left[T(r)\right] \Delta S$.

In the optically thin (ADAF) parts, 
radiative cooling is inefficient due to the very low density.
As a result, accreting matter falls into a central object without losing
its internal energy via radiation. 
Hence, the disk can be significantly hotter with $T_{\rm e} \gsim 10^9$K,
producing high energy (X-$\gamma$ ray) photons.
We use the spectrum calculated by Manmoto, Mineshige, \& Kusunose (1997)
and derive luminosity, $\Delta L(\nu,r)= 2\epsilon_{\nu}(r) \Delta S$,
where $\epsilon_{\nu}$ is emissivity.
Included are synchrotron, bremsstrahlung, and inverse
Comptonization of soft photons created by the former two processes
(see Narayan \& Yi 1995 for details).
Photon energy is thus spread over a large frequency range;
from radio (due to synchrotron) to hard X-$\gamma$ rays (via inverse Compton).
Unlike the optically thick case, radiative cooling is no longer
balanced with viscous heating (and with gravitational energy release).
Consequently, emission from regions within 
a few tens to hundreds of $r_{\rm g}$ 
from the center contributes to most of the entire spectrum 
(see Yonehara et al. 1997).

In the present study, we assign the viscosity parameter to be $\alpha=0.55$.
The mass of the central black hole is $M = 10^8 M_{\odot}$
and the mass-flow rates are $\dot{M} =8 \times 10^{24}$ g s$^{-1}$.
We focus our discussion on the relative changes of the radiation flux and 
variation timescales, and are not concerned with the absolute flux.
It is easy to confirm that the parameters associated with 
the accretion disk only affect non-microlensed flux, 
and do not affect the timescale of microlensing events.
Although the only influential parameter is $v_{\rm t}$, 
it can be estimated by analysis of the optical light curves, 
since the theory of the optically thick disk is relatively well established.
For simplicity, we assume that the accretion disk 
is face-on to the observer ($i=0$).
If we change the inclination angle (i.e., $i \ne 0$), 
the timescale will be shortened by $\cos i$ because of 
the apparently smaller size of the emitting regions,   
but the basic properties are not altered significantly. 
Any how, a quasar is believed to be more or less close to a face-on view.  
The inclusion of the relativistic effects 
(e.g., Doppler shifts by the disk rotation, 
beaming, or gravitational redshifts)  
in the disk model could reveal the detailed disk structure 
in the real vicinity of a central black hole, which will be discussed 
in a future paper.

\subsection{Spectral Variation}

First, we calculate the predicted spectra for the cases 
with the angular separation ($d$) between the caustic and the source center 
normalized with $\theta_{\rm E}$ at 
$d=0.1$, $0.01$, $0.0$, $-0.01$, and $-0.1$, respectively.
The resultant spectral modifications by the lens are shown in figure 2.
For the optically-thick parts, not only disk brightening  
but also substantial spectral deformation are expected 
for $d \le 0.1$.
The spectral shapes critically depend on whether the emitting region is 
inside the caustic or not, since
a large amplification is expected only when
the emitting regions are inside (see Eq.~\ref{eq:ampli}).  
Consequently, totally distinct spectra are expected 
for the cases with $d=0.01$ and $d=-0.01$,
unlike the single-lens calculations (Yonehara et al. 1997).
In the former case,
higher-energy parts (i.e., {\it U}-band) as well as lower-energy parts 
({\it I}-band) are effectively amplified, 
giving rise to a sharply modified spectrum.
In the latter, in contrast, 
the higher-energy part is not substantially amplified,
since the hotter parts are outside the caustic.  
This produces a relatively flat spectrum.  
As a result, frequency-dependent, time asymmetric
microlensing light curves are produced (see \S 3.3).

For the optically thin parts, on the other hand, there are not 
large spectral modifications produced by microlensing; 
the total flux is amplified more strongly over the entire energy range,
although spectral modifications are bit smaller in the X-ray range.
This is because in ADAFs the large emissivity is achieved only 
at several tens to hundreds of $r_{\rm g}$.  
Therefore, timescales for X-ray variation are somewhat shorter.

\subsection{Light Curves}

Such unique spectral effects lead to interesting behaviors in
the microlensing light curves.  We continuously change the
angular separation between the caustic and the center of the 
accretion disk ($d$) according to $d(t) = v_{\rm t}t/D_{\rm ol}\theta_{\rm E}$ 
and calculate the flux at each frequency, assuming optical-UV flux coming from 
the optically thick parts (standard disk model) and 
X-rays from the optically thin parts (ADAF).
Figure 3 shows the light curves of the microlensing events for the
optically thick and thin parts.
There are two interesting noticeable features. 
First, the light curves of the optically thick parts show strong 
frequency dependence in the sense 
that higher-energy radiation (i.e., {\it U}-band) shows more rapid changes 
than lower-energy ({\it K}-band).
Second, the shapes of the light curves are different between the two models;
the changes are gradual in the optically thick parts, 
on a timescale of a few tens of days, while the variation timescales are 
somewhat shorter for the optically thin parts.
This feature reflects the different emissivity distributions of 
the two disk models in the radial directions.
That is, we can infer the radial disk structure 
(i.e., radial dependence of emission properties) from 
the microlensing light curves at different bands.
For this purpose, simultaneous monitoring by ground-based photometry is 
indispensable; namely, we can obtain a good guess for the model parameters, 
such as the transverse velocity, 
by using the standard-disk model for optical emission. 
We can then determine the distance between the caustic and the disk center 
as a function of the observing time. 
On the basis of these data, we can clarify the radial dependence of 
X-ray emission properties. Since the standard-disk model is 
fairly well established, 
and since it predicts that the effective temperature varies as $r^{-3/4}$, 
irrespective of the magnitude of viscosity, 
it is easy to tell the transverse velocity from the time changes of 
estimated effective temperature of the region brightened by a caustic.

In summary, we have studied microlensing of a quasar disk 
by a caustic crossing using realistic models of an accretion disk. 
As a result, we have obtained the basic properties of microlens 
light curves of a quasar like Q2237+0305. 
Future calculations, which incorporate
the effects of caustic curvature and caustic networks,
will tell more precisely how the amplitudes are determined.  

\acknowledgements

The authors would like to express their thanks to 
Joachim Wambsgan{\ss}  for valuable suggestions,
and acknowledge Matt Malkan for helpful comments.
This work was supported in part by the Japan-US Cooperative
Research Program which is founded by the Japan Society for
the Promotion of Science and the US National Science
Foundation, and by the Grants-in Aid of the
Ministry of Education, Science, and Culture of Japan
(08640329, 09223212, SM).

\clearpage

\begin{figure}
\caption{Schematic view of `caustic' crossing during a microlens event.}
\end{figure}

\begin{figure}
\caption{Microlensed and unmicrolensed spectra of 
the optically thick region (standard type disk model: the left panel) 
and the optically-thin one (ADAF model: the right panel), respectively.
In each panel, 
the spectra of an unlensed disk (solid curve)
and those of a lensed disk for the cases with angular separation of
$d=0.1$ (dotted curve), $0.01$ (dashed curve), 
$0.00$ (long dashed curve), $-0.01$ (dot-short dashed curve), and 
$-0.1$ (dot-long dashed curve), are depicted.}
\end{figure}

\begin{figure}
\caption{Microlensing light curves of an optically thick region (upper) 
for the {\it U}-band ($3650 {\rm \AA}$; solid curve),
the {\it V}-band ($5500 {\rm \AA}$; dotted curve,
the {\it I}-band ($8800 {\rm \AA}$; dashed curve),
the {\it K}-band ($22000 {\rm \AA}$; long-dashed curve), and
of an optically thin region (lower) for 2 keV X-rays (by the solid curve), 
respectively.
In each curve, relative flux differences with respect to the values
in the absence of a microlensing are plotted.
The same light curves but on extended time scales are inserted
in the upper right box of each panel.
The mass of a lens object is assumed to be $1.0 M_{\odot}$.}
\end{figure}

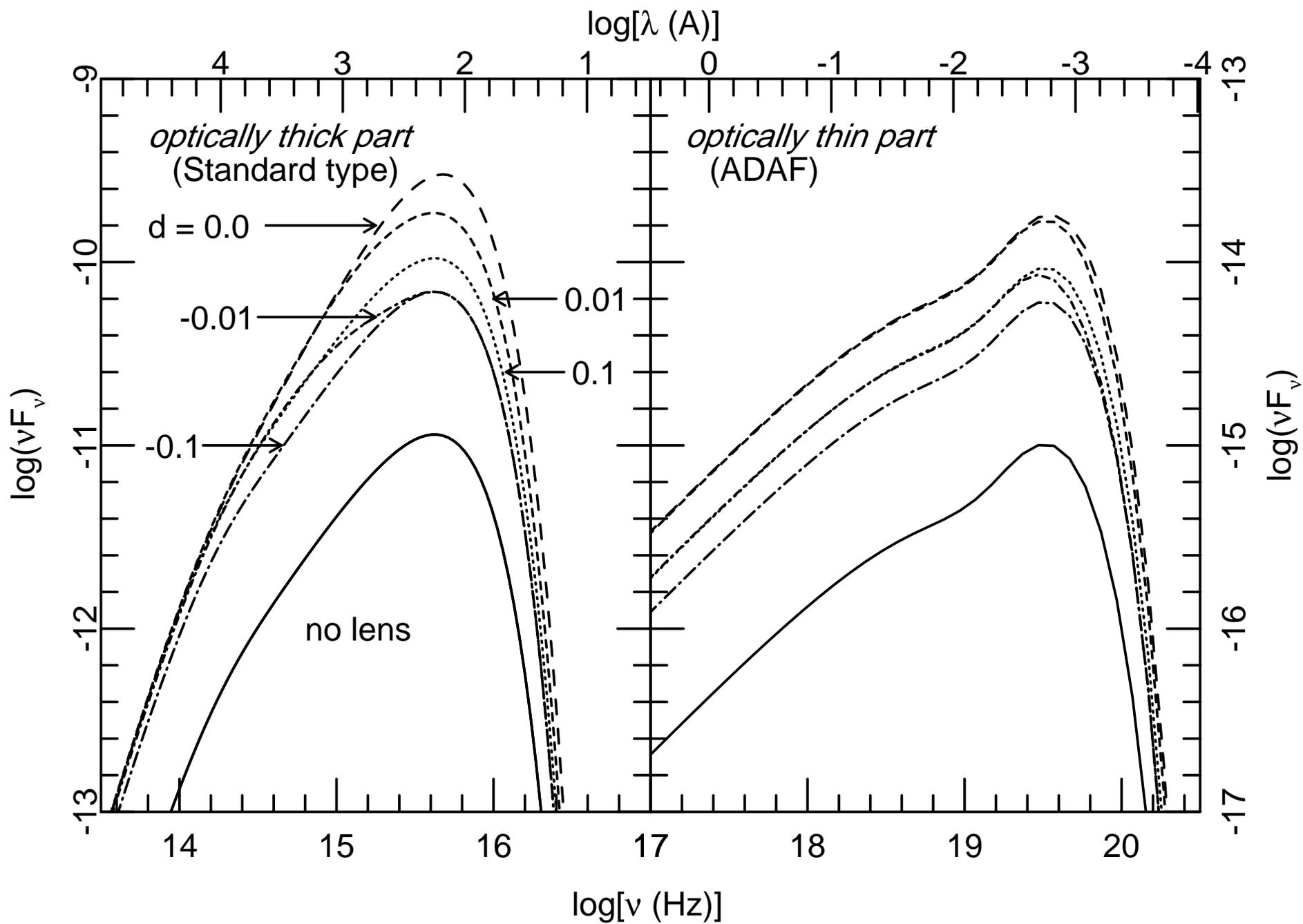
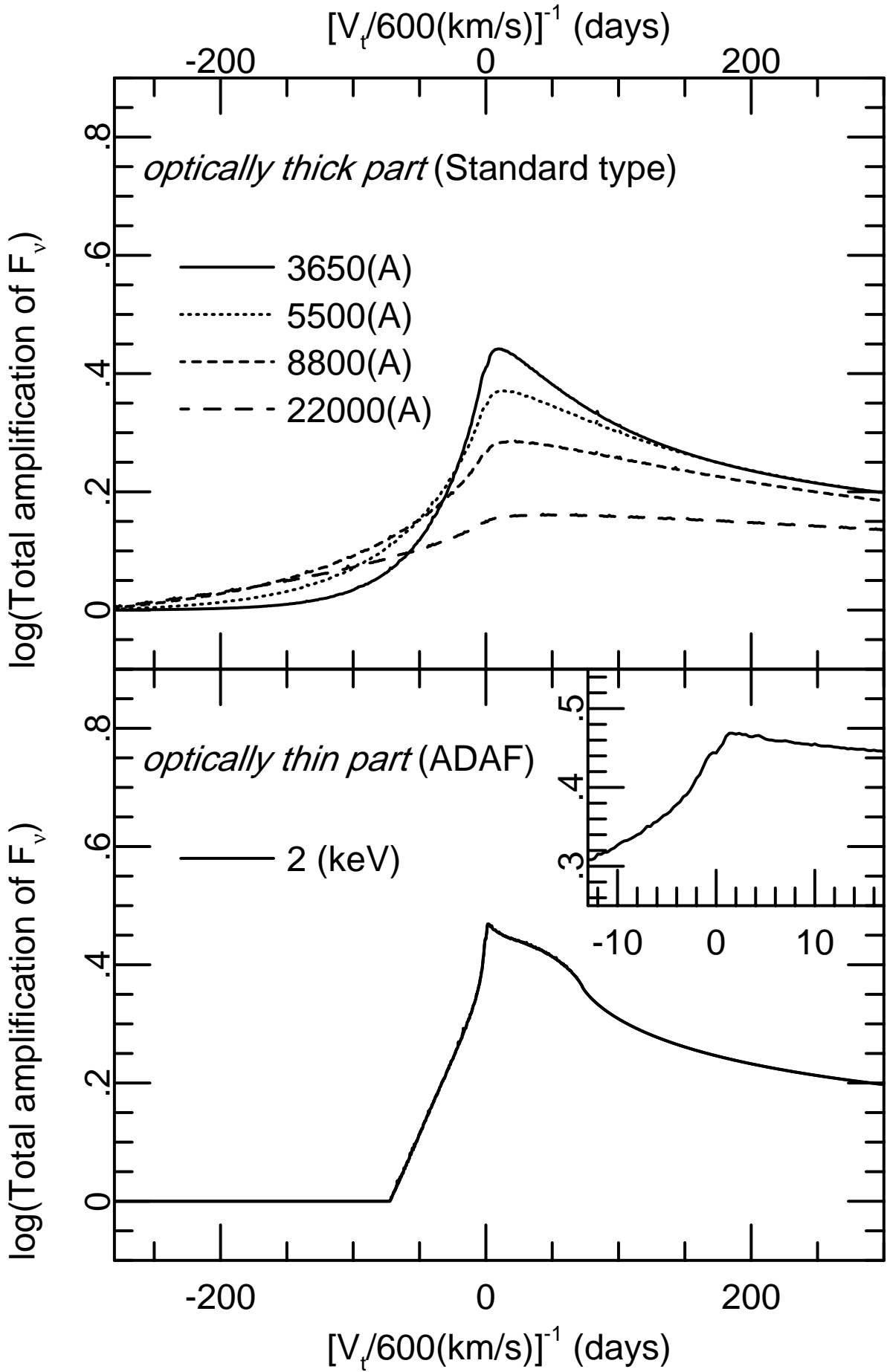
\end{document}